\documentclass[%
 reprint,
 amsmath,amssymb,superscriptaddress,
 aps,
 prl 
]{revtex4-2}

\usepackage{graphicx}
\usepackage{float}
\usepackage{physics}
\usepackage[dvipsnames]{xcolor}

\usepackage[normalem]{ulem}

\newcommand{\fm}{\, {\rm fm}}
\newcommand{\MeV}{\, {\rm MeV}}

\begin{document}

\title{Continuum extrapolated high order baryon fluctuations}

\author{Szabolcs Bors\'anyi}
\affiliation{Department of Physics, Wuppertal University, Gaussstr.  20, D-42119, Wuppertal, Germany}

\author{Zolt\'an Fodor}
\affiliation{Department of Physics, Wuppertal University, Gaussstr.  20, D-42119, Wuppertal, Germany}
\affiliation{Pennsylvania State University, Department of Physics, State College, PA 16801, USA}
\affiliation{Institute  for Theoretical Physics, ELTE E\"otv\"os Lor\' and University, P\'azm\'any P. s\'et\'any 1/A, H-1117 Budapest, Hungary}
\affiliation{J\"ulich Supercomputing Centre, Forschungszentrum J\"ulich, D-52425 J\"ulich, Germany}
\affiliation{Physics Department, UCSD, San Diego, CA 92093, USA}

\author{Jana N. Guenther}
\affiliation{Department of Physics, Wuppertal University, Gaussstr.  20, D-42119, Wuppertal, Germany}

\author{S\'andor D. Katz}
\affiliation{Institute  for Theoretical Physics, ELTE E\"otv\"os Lor\' and University, P\'azm\'any P. s\'et\'any 1/A, H-1117 Budapest, Hungary}

\author{Paolo Parotto}
\affiliation{Pennsylvania State University, Department of Physics, State College, PA 16801, USA}
\affiliation{Dipartimento di Fisica, Universit\`a di Torino and INFN Torino, Via P. Giuria 1, I-10125 Torino, Italy}

\author{Attila P\'asztor}
\affiliation{Institute  for Theoretical Physics, ELTE E\"otv\"os Lor\' and University, P\'azm\'any P. s\'et\'any 1/A, H-1117 Budapest, Hungary}
\affiliation{HUN-REN-ELTE Theoretical Physics Research Group, P\'azm\'any P\'eter s\'et\'any 1/A,1117 Budapest, Hungary}

\author{D\'avid Peszny\'ak}
\affiliation{Institute  for Theoretical Physics, ELTE E\"otv\"os Lor\' and University, P\'azm\'any P. s\'et\'any 1/A, H-1117 Budapest, Hungary}
\affiliation{Department of Computational Sciences, Wigner Research Centre for Physics,
Konkoly-Thege Mikl\'os utca 29-33, H-1121 Budapest, Hungary}

\author{K\'alm\'an K. Szab\'o}
\affiliation{Department of Physics, Wuppertal University, Gaussstr.  20, D-42119, Wuppertal, Germany}
\affiliation{J\"ulich Supercomputing Centre, Forschungszentrum J\"ulich, D-52425 J\"ulich, Germany}

\author{Chik Him Wong}
\affiliation{Department of Physics, Wuppertal University, Gaussstr.  20, D-42119, Wuppertal, Germany}

\date{\today}

\begin{abstract}
Fluctuations play a key role in 
the study 
of QCD phases. Lattice QCD is a valuable tool 
to calculate them, but going to high orders is challenging. 
Up to the fourth order, continuum results are available since 2015. We present the first continuum results for sixth 
order baryon fluctuations for temperatures between $T=130 - 200 \MeV$,  and  
eighth order at $T=145 \MeV$ in a fixed volume. We 
show that for $T \leq 145 \MeV$, relevant for criticality search, finite volume effects are under control. 
Our results are in sharp contrast with well known results in 
the literature obtained at finite lattice spacing.
\end{abstract}

\maketitle

% Heavy ion, phase diagram
\emph{1. Introduction: fluctuations and QCD phases:} 
The main goal of the heavy ion program of many accelerator facilities (e.g., 
at LHC, RHIC or the upcoming CBM/FAIR) is to create new phases of matter 
and explore their properties under extreme conditions. 
Several experimental programs (such as the Beam Energy Scan program at 
RHIC~\cite{STAR:2017sal, STAR:2020tga}) are designed to search for a 
hypothetical critical end point in the temperature-baryon density phase 
diagram.
% Fluctuations are important
Some of the most important observables in this quest are fluctuations of 
conserved charges. In the grand canonical ensemble, they are derivatives 
of the pressure with respect to the chemical potentials coupled to the 
charges. In this work, we will calculate such fluctuation observables at 
zero baryochemical potential.

% -- Equation of state\\
The physics applications of fluctuation observables are numerous. 
First, the equation of state of the hot-and-dense quark gluon plasma is one 
of the main inputs of lattice QCD to the phenomenology of heavy ion physics.
Fluctuations at zero baryochemical potential $\mu_B$ are the basis for 
extrapolations of the QCD equation of state to non-zero $\mu_B$, both
by means of a truncated Taylor expansion~\cite{Bollweg:2022fqq}, as well as
via different resummations of the Taylor 
series~\cite{Borsanyi:2021sxv,Borsanyi:2022qlh,Bollweg:2022rps}.

%-- CEP signatures \\
Second, fluctuation observables are sensitive to criticality. While 
first-principle lattice simulations have shown that the chiral transition 
is a crossover at zero baryon density~\cite{Aoki:2006we}, at larger baryon 
densities several model calculations predict a critical endpoint in the 3D 
Ising universality class~\cite{Kovacs:2016juc, Critelli:2017oub, Isserstedt:2019pgx, Fu:2021oaw}, where the crossover line 
becomes a line of 
first order transitions. One of the 
proposed experimental signatures of 
such a critical endpoint is a 
non-monotonic behavior of the fourth-to-
second order baryon number fluctuations 
as a function of 
$\mu_B$~\cite{Stephanov:2011pb, Mroczek:2020rpm}. The extrapolation of this 
ratio to $\mu_B>0$ is possible, if a sufficient number of Taylor coefficients (fluctuations) are 
available at $\mu_B=0$. Thus, fluctuations at $\mu_B=0$ are also 
important for the quest to find the critical endpoint. An important 
baseline in this search is given by the hadron resonance gas (HRG) model, a 
non-critical model that describes thermodynamics below the chiral 
transition at $\mu_B=0$ remarkably well. 
A reasonable minimum 
criterium for 
criticality searches, then, is the presence of solid deviations between HRG 
predictions and equilibrium QCD. 

%-- O(4) behaviour \\
%-- Criticality in theory \\
%~~ -- Radius of convergergence \\
%~~ -- LY Zeros\\
A different type of criticality - in the O(4) universality class - is also 
expected to be present in QCD, related 
to chiral symmetry restoration, 
one of the most important concepts
in heavy ion physics. In the limit 
of zero light quark masses, the $SU(2) \times SU(2)$ chiral symmetry 
becomes exact, and the crossover transition occurring at physical masses is 
expected to be replaced by a genuine second order 
transition ~\cite{Pisarski:1983ms, HotQCD:2019xnw, Kotov:2021rah}. 
Through the presence of a single scaling variable (a combination of the 
quark masses, the temperature and the chemical potential), O(4) criticality 
has imprints on the temperature dependence of higher baryon number 
fluctuations at $\mu_B=0$ up to physical values of the quark masses (for a 
model calculation, 
see e.g. Ref.~\cite{Almasi:2017bhq}). 
The presence of 
such an O(4) scaling regime is likely 
the reason why lattice QCD calculation 
see no sharpening or strengthening of 
the crossover transition for small 
chemical 
potentials~\cite{Bonati:2015bha, Borsanyi:2020fev, Borsanyi:2021sxv}.
The complicated interplay of O(4) and 
Ising criticality motivates more 
involved theoretical approaches to the 
phase diagram, based on Lee-Yang 
zeroes~\cite{Yang:1952be,Lee:1952ig}, which 
have gained popularity in 
recent 
years~\cite{Giordano:2019slo, Mukherjee:2019eou, Giordano:2019gev, Basar:2021hdf, Dimopoulos:2021vrk}. 
These approaches, too, require reliable determinations of the corresponding 
fluctuations.

Third, fluctuation observables are very sensitive to the degrees of freedom 
of a thermodynamic system. This fact has been used before to argue, e.g., 
the existence of further (yet undiscovered) 
resonances~\cite{Majumder:2010ik, Bazavov:2014xya, Alba:2017mqu} in the
hadron spectrum, which was later confirmed by 
experiment~\cite{ParticleDataGroup:2022pth}.

%-- Freeze-out physics\\
%~~ -- extrapol of fluct\\
Finally, fluctuation observables are central in the study of chemical 
freezeout in heavy ion collisions. This is an especially interesting avenue 
of research, since it has the potential of allowing a direct comparison 
between first-principle QCD predictions and experimental data. While the 
comparison itself has many 
caveats, due to experimental effects~\cite{Bzdak:2012an, Braun-Munzinger:2020jbk, Vovchenko:2020tsr}, it 
is obviously worth pursuing.

%Situation on the lattice \\
\emph{2. Current lattice estimates:} 
Conserved charge fluctuations have been a focus of lattice simulations for 
well over a decade now. Like any other
observable, a reliable calculation of
these requires a continuum limit extrapolation,
via simulations using smaller-and-smaller
lattice spacings. Up to second order, they are known in the continuum 
since 2012~\cite{Borsanyi:2012cr}. Fourth order fluctuations in the baryon 
number and strangeness were first continuum extrapolated in 
2015~\cite{Bellwied:2015lba}. In that case, a large temperature range was 
considered, showing good agreement with the hadron resonance gas model at 
low temperatures, as well as good agreement with perturbative 
calculations~\cite{Mogliacci:2013mca, Haque:2013sja} at high temperatures. 
Since then, calculations of the fourth order coefficients were pushed to 
very high precision~\cite{Bazavov:2020bjn}. Thus, up to fourth order, the 
derivatives in full QCD are known
accurately, with the exception of electric 
charge fluctuations, which suffer from large cut-off effects that make 
continuum extrapolations difficult~\cite{Bellwied:2015lba}.

At the sixth and eighth orders, the statistics requirements for the direct 
determination of the coefficients dramatically blow up, thus, a continuum 
extrapolation of these coefficients has never been attempted. 
Because of the high statistics required, all available results on higher 
fluctuations employ the computationally cheapest discretization: staggered 
fermions. These suffer from a lattice artefact called taste 
breaking~\cite{Sharpe:2006re, MILC:2009mpl}, whose effect is to 
strongly 
distort the meson spectrum at a finite 
spacing. One might think that, since 
baryon number fluctuations directly couple 
to baryons only, and not mesons, 
this issue is of no great importance 
here. Unfortunately, this is only the 
case at the level of the baryon 
number $B=1$ sector of the Hilbert space. 
Two-baryon interactions are mediated by 
mesons, and thus, taste breaking distorting
meson physics
can have relevant effects on the description 
of baryon dense matter (or, 
equivalently, on high order fluctuations at $\mu_B=0$). 
In fact, the higher the order of 
the fluctuations, the larger the effect of 
multi-baryon interactions on 
them. Hence, these fluctuations could 
be particularly sensitive to taste 
breaking effects. On a more basic 
level, higher order 
$\mu_B$-derivatives probe physics at larger $\mu_B$, which at a finite 
spacing will be closer to the cut-off scale of $1/a$ ($a$ being the lattice 
spacing), making cut-off effects potentially important.

The statistics required for the calculation of higher order fluctuations 
can be drastically reduced by introducing a purely imginary chemical 
potential, calculating lower order fluctuations, and fitting their 
functional dependence on the imaginary chemical potential. The price for 
this reduction in statistics requirements is that assumptions have to be 
made on the functional form of the lower order fluctuations, leading to 
hard-to-control systematic errors.

So far, three collaborations have presented results up to the eighth order 
with improved lattice actions. In chronological order: First, the Pisa 
group presented results on lattices with 6 timeslices of 2stout improved 
fermions~\cite{Aoki:2006we, Borsanyi:2010bp} in Ref.~\cite{DElia:2016jqh}. 
Second, the Wuppertal-Budapest collaboration presented results with 12
timeslices of 4stout improved fermions~\cite{Borsanyi:2016ksw} in 
Ref.~\cite{Borsanyi:2018grb}. These two calculations took advantage of 
simulations at imaginary chemical potential. Finally, the HotQCD 
collaboration presented results with 8 timeslices of HISQ 
fermions~\cite{Follana:2006rc}, using a direct determination at $\mu_B=0$ 
(i.e., without imaginary $\mu_B$ simulations) in 
Refs.~\cite{Bazavov:2020bjn, Bollweg:2022rps}. 
% \pp{Why do you write Taylor? Also later. Direct determination is independent of Taylor.}
For the latter calculation, two orders of magnitude more statistics were 
collected, compared to the previous two. It is a testament to the 
efficacy of the imaginary chemical potential method, then, that the error 
bars on the imaginary chemical potential calculations of the sixth and 
eighth order coefficients are substantially smaller. 

At the current level of precision, 
discrepancies emerge between the calculations.
In particular, the results based on the 
imaginary chemical potential method 
are in good agreement with the hadron 
resonance gas model for low 
temperatures. On the other hand, the 
direct calculation shows significant 
deviations for both observables even 
at the lowest temperature considered. 
To shed light on QCD criticality, this discrepancy has to be resolved.

\begin{figure*}[t!]
    \begin{center}
        \includegraphics[width=0.325\linewidth]{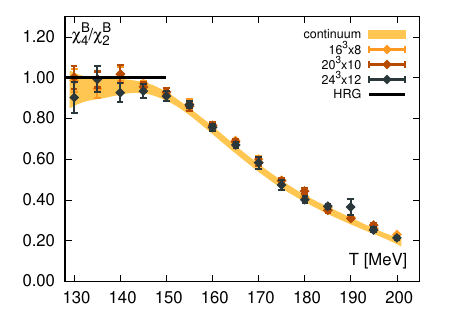}
        \includegraphics[width=0.325\linewidth]{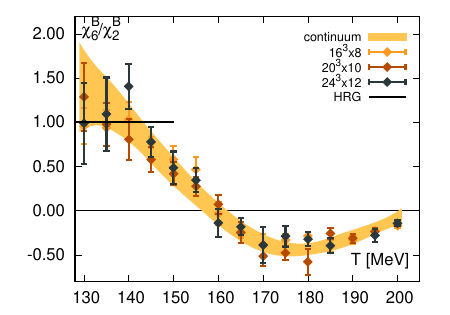}
        \includegraphics[width=0.325\linewidth]{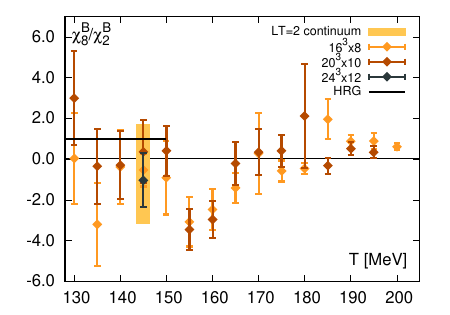} 
    \end{center}
    \vspace{-0.6cm}
    \caption{Our lattice results for the ratios $\chi^B_4/\chi^B_2$ (left), $\chi^B_6/\chi^B_2$ (center), $\chi^B_8/\chi^B_2$ (right).  For the first two, the continuum extrapolation is shown as a yellow band. HRG model predictions are shown as solid black lines in all cases.
    \label{fig:chi_ratios}        
}
\end{figure*}

In addition to the different extraction method for the higher order 
coefficients (direct at $\mu_B=0$ vs indirect from $\mu_B^2 \leq 0$), a potentially 
more significant difference between the two types of calculation lies in how
the chemical potential is defined. 
Due to the extreme cost of the direct 
method, for fluctuations of order six and higher, the chemical potential in 
that case was introduced via the so-called 
linear prescription. In the other two 
cases, the exponential definition was used.
Derivatives with respect to the chemical potential can be shown to be UV 
finite by virtue of a $U(1)$ symmetry if the chemical potential is 
introduced like a constant imaginary gauge field~\cite{Hasenfratz:1983ba}. 
This is the exponential definition. If, instead, a naive linear definition 
is employed, power-law UV divergences appear in the free energy 
already in the case of free quarks. By taking enough derivatives with 
respect to the chemical potential, such power-law divergences disappear. 
However, this is not the case for logarithmic divergences, the absence of 
which for the naive linear definition is 
not proven for the interacting theory. 
Thus, although the linear definition is 
computationally cheaper, care 
should be taken in considering these results. 

The 
linear definition also breaks the exact Roberge-Weiss 
periodicity~\cite{Roberge:1986mm} of the
partition function. Even if one assumes that
there are no problems with 
logarithmic divergences,
the loss of Roberge-Weiss periodicity with 
the linear definition can potentially lead to
large cut-off effects, since it effectively
means that at a finite spacing, in contrast
to the continuum, the free energy
gets contributions from Hilbert subpsaces
not only at integer, but also at 
non-integer values of the baryon number, which
at the very least, is a non-physical feature 
at finite spacing.

\emph{3. Lattice calculation of fluctuations up to eighth order}

In this letter, we present the first continuum results for baryon 
fluctuations up to the sixth order for temperatures between 
$T = 130 - 200 \MeV$, and up to the eighth order for a temperature of 
$T=145 \MeV$. Continuum extrapolation is made possible by the introduction 
of a new discretization, which we call the 4HEX action, that strongly 
suppresses taste breaking effects compared to all available actions in the 
literature. 
Although more costly,  we pursue a direct determination at $\mu_B=0$, in order to avoid possible systematic effects due to 
a choice of fit ansatz, necessary for the imaginary chemical potential method. 
Moreover, in order to avoid possible issues with the introduction of the 
chemical potential, we employ the exponential definition to all orders. 
Due to the extreme statistics cost of the direct method, this endeavour is 
only feasible in a volume that is smaller than what is typically used in 
the field, with an aspect ratio $LT=2$. Thanks to the availability in the 
literature of the aforementioned results at finite lattice spacing, but 
with larger volume, we are able to show that below $T=145 \MeV$, finite 
volume effects in our results are under control. Note that this is the 
relevant temperature range for the search for the elusive critical endpoint of QCD. 

% intro ]]]

%\section{\label{sec:continuum}Continuum results} %[[[

%-- 4HEX \\
The novel lattice action we use for this thermodynamics study, 4HEX, is 
based on rooted staggered fermions with 4 steps of HEX 
smearing~\cite{Capitani:2006ni} with physical quark masses, and the DBW2 
gauge action~\cite{QCD-TARO:1998nbk}. 
This lattice action benefits from dramatically reduced taste breaking 
effects, compared to all other actions used in the literature. 
We simulate $16^3 \times 8$, $20^2 \times 10$ and $24^3 \times 12$ lattices 
to obtain a well-controlled continuum extrapolation. 
Details on the 4HEX action, the scale setting procedure, and the systematic error 
estimation can be found in the supplemental material.

We calculate fluctuations of the baryon number at zero strangeness 
chemical potential:
\begin{equation}
    \chi^B_n \equiv \left( \frac{\partial ^n (p/T^4)}{\partial (\mu_B/T) ^n} \right)_{\mu_S=0} \rm.
\end{equation}

We also include results on the strangeness neutral line 
$n_s \equiv 0$ in the Supplemental Material, which lead to similar 
conclusions as in the $\mu_S=0$ case.

%-- Reduced matrix formalism \\
%-- Exponential definition in all orders\\
We use the exponential definition of the chemical potential at all orders 
in $\mu_B$ on all our lattices. For the $N_\tau =8, 10$ lattices, we 
use the reduced matrix formalism to calculate the fluctuations, in the same 
way as we did in Refs.~\cite{Borsanyi:2022soo, Borsanyi:2023tdp}. For the 
$N_\tau=12$ lattice, we use the standard random source 
method~\cite{Allton:2002zi}. 

We show our continuum extrapolated results for $\chi^B_4/\chi^B_2$ (left) 
and $\chi^B_6/\chi^B_2$ (center), together with the corresponding finite 
lattice spacing results in Fig.~\ref{fig:chi_ratios}. The continuum results are obtained together with a spline fit of the temperature dependence. The exact procedure is described in the Supplemental Material. The bands include statistical and systematic uncertainties, consisting of different scale settings and different spline fits of the data. 
The covered temperature
range is $130 \MeV \leq T \leq 200 \MeV$. Also shown are the results on
the $N_\tau=8, 10$ lattices for $\chi^B_8/\chi^B_2$ (right). For this 
observable, we also include the continuum extrapolation at a single 
temperature of $T=145 \MeV$. 
Hadron resonance gas predictions are shown, and  they equal 1 in all cases 
independently from the temperature and the hadron spectrum used.

From Fig.~\ref{fig:chi_ratios}, it is apparent how small the cut-off 
effects of the 4HEX action are, as is the fact that, for $T<145 \MeV$, the 
fluctuations in continuum QCD are in very good agreement with the HRG 
results.

\emph{4. Comparisons with the literature:}

\begin{figure}[t]
    \begin{center}
        \includegraphics[width=\linewidth]{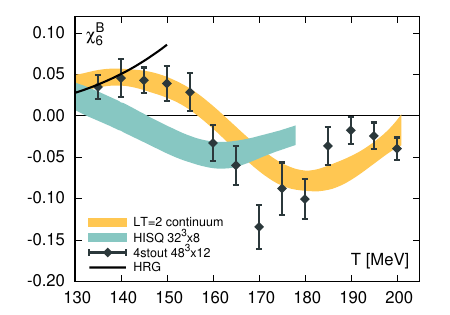}  
        \includegraphics[width=\linewidth]{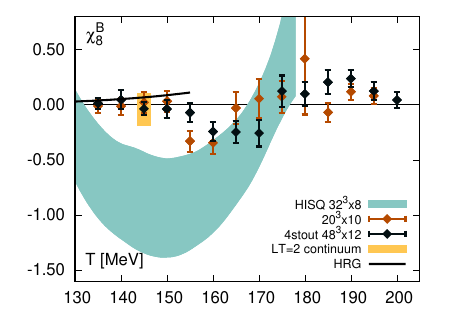}   
    \end{center}
    \vspace{-0.6cm}
    \caption{
    \label{fig:chis} Our results for sixth (top) and eighth (bottom) order baryon number susceptibilities. The HISQ (splined) results are shown as light blue bands, and the 4stout results are shown are black points. Our new results are shown as a yellow band for the continuum extrapolation of $\chi_6^B$ and the continuum extrapolation of $\chi_8^B$ (at $T=145\MeV$), and as brown points for $\chi_8^B$ on the $20^3\times 10$ lattice. HRG model predictions are shown as solid black lines. 
}
\end{figure}

Being our results on the sixth and eighth order fluctuations the first ever 
continuum extrapolated, we proceed to compare them to previous results 
from the literature, with an aspect ratio $LT=4$. 
In the top panel of Fig.~\ref{fig:chis}, we show the continuum $\chi^B_6$ (yellow band) 
alongside previous results by the Wuppertal-Budapest collaboration obtained 
with the imaginary chemical potential method using 12 timeslices of 4stout 
improved staggered fermions~\cite{Borsanyi:2018grb} (black points), and 
results by the HotQCD collaboration obtained with the direct method and 8 
time slices of HISQ fermions~\cite{Bollweg:2022rps} (green band). The 
latter are not direct data, but rather come from a spline interpolation of 
the direct data. The comparison is fair, since we also use a spline 
interpolation on our data, to allow for systematic error estimation on the 
continuum limit. The role of the 
spline interpolation in thermodynamics
studies is to i) reduce the 
random noise on the direct data, by utilizing
continuity in $T$ and to ii) allow one to 
take temperature derivatives, which are 
needed for the calculation of 
certain thermodynamic 
quantities, such as the speed of sound. The HotQCD collaboration has also used the splined
version of their results on high order fluctuations in determining 
phenomenological quantities (see, e.g., Ref.~\cite{Bollweg:2022rps}). 
Looking at the three results in comparison, a very simple interpretation 
emerges. First, at temperatures below $145 \MeV$, our results at $N_\tau=12$
agree with our new continuum extrapolated results, even though the volume 
is different in the two simulations. On the other hand, the $N_\tau=8$ 
HotQCD result does not agree with our old $N_\tau=12$ results, even though 
the physical volumes are the same. This means that, at low temperatures, 
finite volume effects are small and cut-off effects are large. It appears 
that these $N_\tau=8$ lattices are too coarse for phenomenological 
applications. The $N_\tau=8$ results show very
significant deviations from the hadron 
resonance gas. For example, it is 
noteworthy that the sign of $d\chi^B_6/dT$ is opposite to the HRG model 
prediction at all these low temperatures. 
Such deviations turned out to be cut-off effects, since the continuum 
extrapolated results show good quantitative agreement with the HRG.

A similar scenario appears in the bottom panel of Fig.~\ref{fig:chis}, where
we compare our new results on $20^3\times 10$ lattices with results from the same 
$LT=4$ simulations -- Wuppertal-Budapest $N_\tau=12$~\cite{Borsanyi:2018grb} (black points) and HotQCD $N_\tau=8$~\cite{Bollweg:2022rps} (green band) -- as shown in the case of $\chi_6^B$. We also include our new continuum extrapolation at $T=145 \MeV$. Besides the 
markedly smaller errors, our results are in good agreement with each other, 
showing small volume dependence especially at lower temperature. Moreover, 
quantitative agreement with HRG model predictions is evident up to 
$T=145 \MeV$. Our continuum result at $T=145 \MeV$ confirms these findings.
As in the previous case, it appears that cut-off effects for $T \leq 145 \MeV$ are too large to
allow for a safe use of results on coarse lattices for phenomenological 
applications. 
Finally, we note that 
the
for $T>145 \MeV$, our old $N_\tau=12$ large 
volume and new $N_\tau=10$ small volume 
results do not agree. In particular, the local
minimum of $\chi^B_8$ is shifted to 
lower temperatures. This is likely a 
finite volume effect, due to the crossover
transition moving to slightly lower $T$ in
a smaller volume.

\emph{5. Discussion:} In this letter we have reported the first continuum 
extrapolated results for high order baryon number fluctuations available in 
the literature. By means of a novel discretization of the QCD action, we 
were able to carry out a continuum extrapolation using lattice with 8,10 and 
12 time slices. We used an aspect ratio of  $LT=2$. We calculated sixth 
order fluctuations in the continuum in a temperature range between 
$T = 130 - 200 \MeV$. We also calculated the eighth order fluctuations 
in the continuum at a single temperature $T=145 \MeV$. 
A comparison of our results with existing results at finite lattice 
spacing and larger physical volumes showed that, in the temperature regime 
relevant for the critical point search, volume effects are well under 
control already for the smaller volume used in our study. In contrast, 
cut-off effects in previous results in the literature where not always 
under control, especially at the lower temperatures relevant for 
constraining the position of the critical endpoint. 
Thus, phenomenological conclusions based on erroneous coefficients ought to 
be reexamined in the near future. These include estimates of the radius of 
convergence and poles of Pad\'e approximants, used to constrain the location 
of the critical endpoint and the convergence of the Taylor expansion for the 
equation of state, used as input for hydrodynamic simulations as well as 
estimates of fluctuation observables at $\mu_B>0$ used to study chemical 
freeze-out. We plan to revisit all of these
points in future publications.

% methods ]]]

% continuum ]]]

\emph{Acknowledgements:} The project was supported by the BMBF Grant
No. 05P21PXFCA. This work is also supported by the
MKW NRW under the funding code NW21-024-A. Further
funding was received from the DFG under the Project
No. 496127839. This work was also supported by the
Hungarian National Research, Development and Innovation
Office, NKFIH Grant No. KKP126769.
This work was also supported by the NKFIH excellence
grant TKP2021{\textunderscore}NKTA{\textunderscore}64. D.P. is supported by the ÚNKP-23-3 New National Excellence Program of the Ministry for Culture and Innovation from the source of the National Research, Development and Innovation Fund.
The authors gratefully acknowledge the Gauss Centre for
Supercomputing e.V. (\url{www.gauss-centre.eu}) for funding
this project by providing computing time on the GCS
Supercomputers Juwels-Booster at Juelich Supercomputer
Centre and HAWK at H\"ochstleistungsrechenzentrum Stuttgart.

\section{Supplemental material}

\subsection*{\label{4HEX}4HEX action}

\begin{figure}
\center
\includegraphics[width=\linewidth]{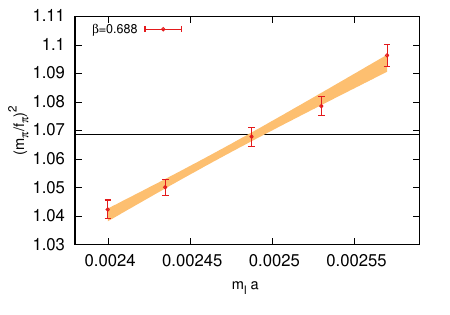}
\includegraphics[width=\linewidth]{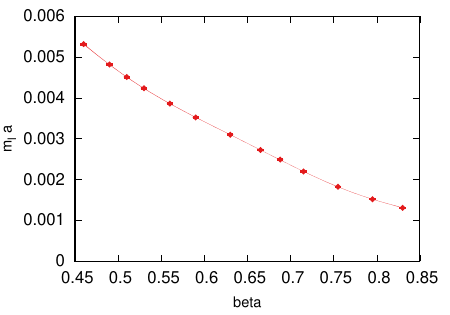}
\includegraphics[width=\linewidth]{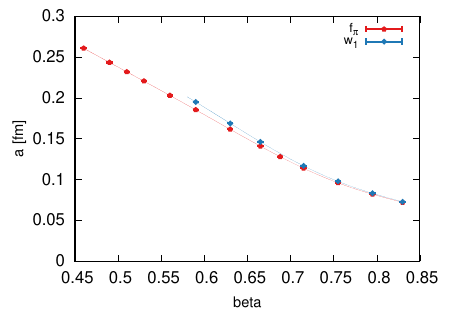}
\caption{Top: Example of quark mass tuning for simulations at 
$\beta=0.688$. The orange band is a linear fit of the data, and the 
black line is the target value of the ratio $m_\pi^2/f_\pi^2$.
Center: Line of constant physics i.e., light quark mass as a function of 
the inverse bare gauge coupling. The points show the direct determination 
from the procedure in the top panel, the band is a polynomial fit as a 
function of $\beta$.
Bottom: lattice spacing as determined through the pion decay constant (red) 
and $w_1$ scale (orange). The band is a polynomial fit as a function of the coupling
$\beta$.}
\label{fig:mass_tuning}
\end{figure}

In this work we employ simulations with $N_f=2+1+1$ staggered 
dynamical quark flavours with physical masses. In the gauge sector 
we use the DBW2 action \cite{QCD-TARO:1998nbk}, and in the fermion 
sector we use links with 4 levels of HEX smearing 
\cite{Capitani:2006ni}. 

Before quantitative results can be extracted, the action must be 
parameterized, meaning that the dependence of the bare quark masses
on the gauge coupling must be tuned, so that the theory is 
renormalized in the parameter range of interest. In our simulations 
we used degenerate up and down quark masses, and fixed the light-
strange mass ratio to $m_s/m_{ud}=27.63$, and the strange-charm mass 
ratio to $m_c/m_s=11.85$. In order to tune the light quark mass, we
used the ratio $m_\pi/f_\pi = 1.0337$, where we used the 
isospin-averaged pion mass $m_\pi=134.8 \MeV$ \cite{Aoki:2013ldr} 
and the pion decay constant $f_\pi = 130.41 \MeV$. In order to 
adopt a proper comparison, we corrected both $m_\pi$ and $f_\pi$ for
finite volume effects using results from two-flavour chiral 
perturbation theory \cite{Colangelo:2003hf}. For each value of the 
coupling, we simulated 3-5 values of the quark masses, then 
extracted the tuned masses via a linear interpolation in 
$m_\pi^2/f_\pi^2$. An example of the mass tuning for $\beta=0.688$ 
is shown in the top panel of Fig.~\ref{fig:mass_tuning}.

\begin{figure}
    \centering
    \includegraphics[width=\linewidth]{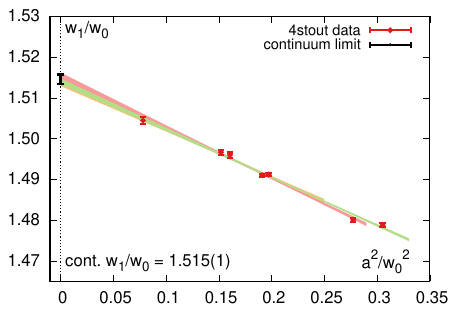}
    \caption{Continuum limit of the ratio $w_1/w_0$ from 4stout simulations.}
    \label{fig:w1w0}
\end{figure}

We parameterized the action in the range $\beta=0.46-0.83$, 
corresponding to lattice spacings in the range $a=0.261-0.072 \fm$. The 
line of constant physics is shown over the whole range in the center panel
of Fig.~\ref{fig:mass_tuning}.
We set the scale with the pion decay constant $f_\pi$, as well as 
with a modified version of the Wilson-flow-based 
$w_0$~\cite{Borsanyi:2012zs}, which we refer to as $w_1$. Both scale settings are shown in the bottom panel of 
Fig.~\ref{fig:mass_tuning}. The definition
for $w_0$ is the following:
\begin{equation}
\left. W (t)\right|_{t=w^2_0} = 0.3 \, \, ,
\end{equation}
where $t$ is the flow time, and $W(t)$ is defined as:
\begin{equation}
W (t) \equiv t \frac{d}{dt} \left\lbrace t^2 \left\langle E(t) \right\rangle \right\rbrace \, \, , 
\end{equation}
where, in turn, $\left\langle E(t) \right\rangle$ is the expectation value 
of the continuum-like action density $G^a_{\mu\nu}(t) G^a_{\mu\nu}(t)/4$.
The quantity $w_1$ is similarly defined as:
\begin{equation}
\left. W (t)\right|_{t=w^2_1} = 0.7 \, \, ,
\end{equation}
and the continuum value of the ratio $w_q/w_0$ is $w_1/w_0 = 1.515(1)$. We 
show the approach to the continuum of this ratio in Fig.~\ref{fig:w1w0}, 
where three linear fits differing by the range considered are shown. The 
final continuum value is obtained combining the different fit results.

\begin{figure}[t!]
    \centering
    \includegraphics[width=\linewidth]{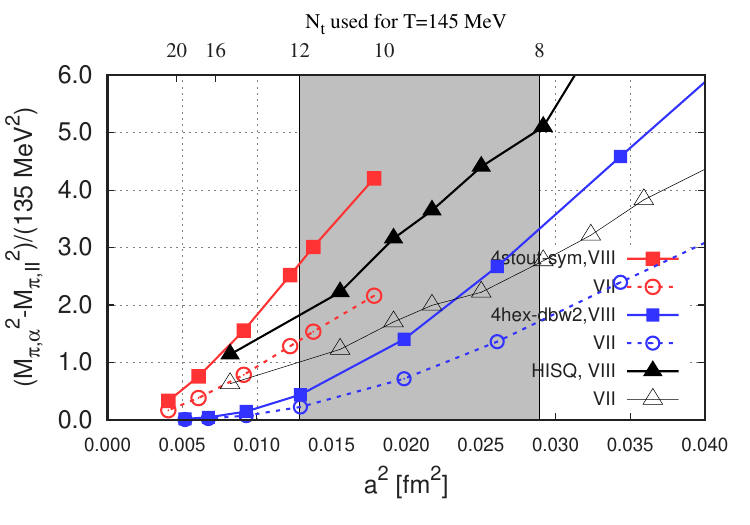}
    \caption{Taste breaking for our previous 4stout action with Symanzik-improved gauge sector (red), HotQCD's HISQ action (black), and our new 4HEX action with DBW2 gauge sector (blue). The improvement in the taste breaking is evident.}
    \label{fig:taste_breaking}
\end{figure}

The main advantage of the new action we are introducing is the 
reduced taste breaking one observes in the pion sector. In 
Fig.~\ref{fig:taste_breaking} we show this as a function of 
the lattice spacing for our previously used 4stout action (red) and 
our new 4HEX action (red). It is clear that the deviation of the 
pion masses from the target value are much smaller, by up to more 
than one order of magnitude for finer lattices. We will see in the 
following that this has crucial consequences on the feasibility of
a continuum extrapolation with accessibly fine lattices.

\subsection*{\label{systematic}Systematic analysis}

In order to obtain the continuum extrapolations shown in 
Figs.~\ref{fig:chi_ratios},~\ref{fig:chis}, we used the following ansatz:
\begin{equation}
    \hat{O}(T, 1/N_\tau^2) = \sum_{i=1}^M \left( \alpha_i + \beta_i \frac{1}{N_\tau^2} \right) s_i(T) \, \, ,
\end{equation}
where the $s_i(T)$ are a set of basis spline functions. In order to take 
into account the ambiguity in the temperature definition, we employ settings
of the scale with both $f_\pi$ and $w_1$, as previously mentioned, and 
include three different sets of node points to define the basis functions
$s_i(T)$. The final results are obtained by combining the $6 = 2\times 3$ 
analyses to construct a histogram. The width of the histogram defines the
systematic error. In the plots we show combined errors, where we assume 
that statistical and systematic errors add up in quadrature. This 
systematic error estimation procedure has been used and described in more 
detail in several of our works, most recently in 
Ref.~\cite{Borsanyi:2018grb}. 

\begin{figure}
    \centering
    \includegraphics[width=\linewidth]{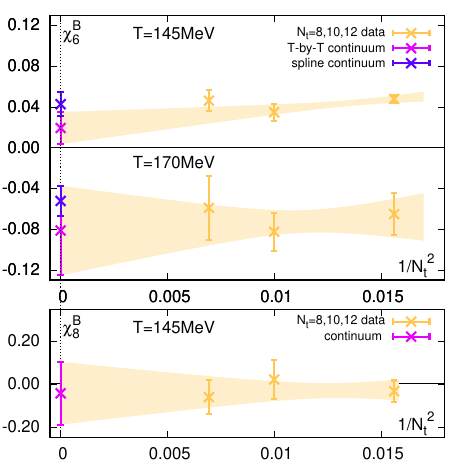}
    \caption{Continuum extrapolation of $\chi_6^B$ (top panel) at $T=145,170 \MeV$ and $\chi_8^B$ (bottom panel) at $T=145 \MeV$. For the different temperatures, the finite $N_\tau$ data are shown along with their continuum extrapolation. In the case of $\chi_6^B$, the results of the combined spline and continuum fit (depicted in Fig.~\ref{fig:chis}), are also shown.}
    \label{fig:chi68cont}
\end{figure}

We show in Fig.~\ref{fig:chi68cont} two example continuum extrapolations 
of $\chi_6^B$, at $T=145,170 \MeV$, and the continuum extrapolation of 
$\chi_8^B$ at $T=145\MeV$. In all cases, one can appreciate the small 
cutoff dependence of these quantities. In the case of $\chi_6^B$, we show 
the continuum extrapolation in pink, together with the result of the 
spline + continuum fit with the ansatz described in the previous 
paragraph. 

\subsection*{\label{sn} Strangeness Neutrality}

\begin{figure*}[t!]
    \begin{center}
        \includegraphics[width=0.325\linewidth]{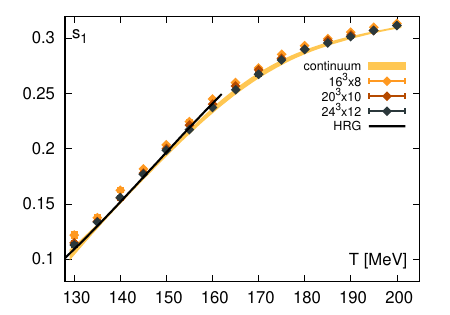}
        \includegraphics[width=0.325\linewidth]{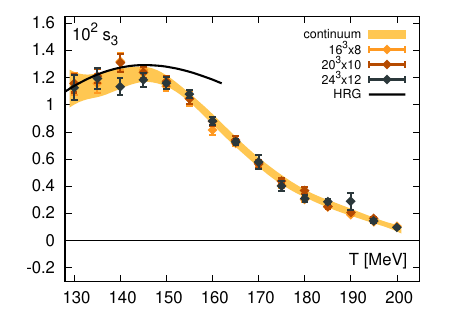}
        \includegraphics[width=0.325\linewidth]{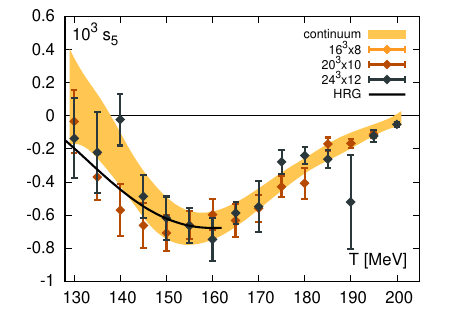} 
    \end{center}
    \vspace{-0.6cm}
    \caption{Our lattice results for the coefficients $s_1$ (left), $s_3$ (center), $s_5$ (right). The continuum extrapolation is shown as a yellow band. HRG model predictions are shown as solid black lines.
    \label{fig:s1s3s5}        
}
\end{figure*}

\begin{figure*}[t!]
    \begin{center}
        \includegraphics[width=0.325\linewidth]{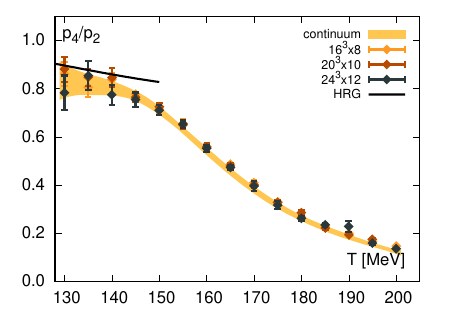}
        \includegraphics[width=0.325\linewidth]{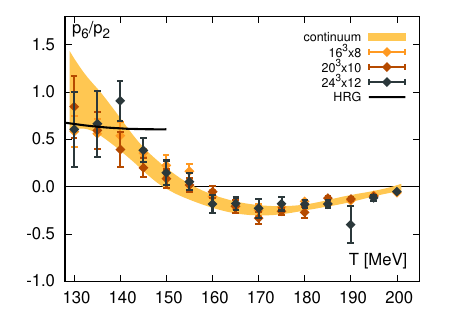}
        \includegraphics[width=0.325\linewidth]{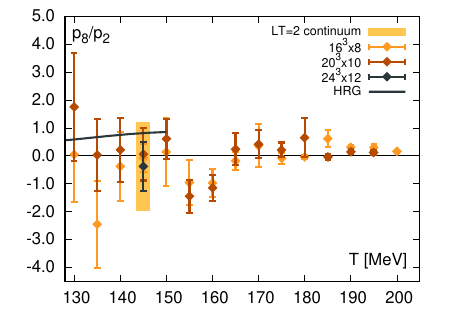} 
    \end{center}
    \vspace{-0.6cm}
    \caption{Our lattice results for the strangeness neutral ratios $p_4/p_2$ (left), $p_6/p_2$ (center), $p_8/p_2$ (right).  For the first two, the continuum extrapolation is shown as a yellow band. HRG model predictions are shown as solid black lines.
    \label{fig:p42p62p82}        
}
\end{figure*}

\begin{figure}
    \begin{center}
        \includegraphics[width=\linewidth]{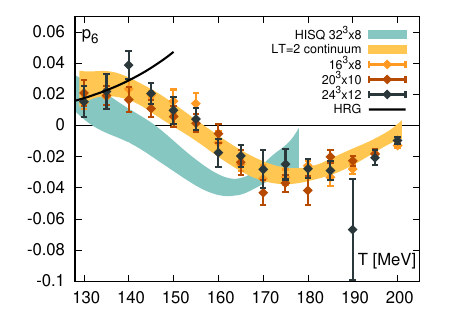}
        \includegraphics[width=\linewidth]{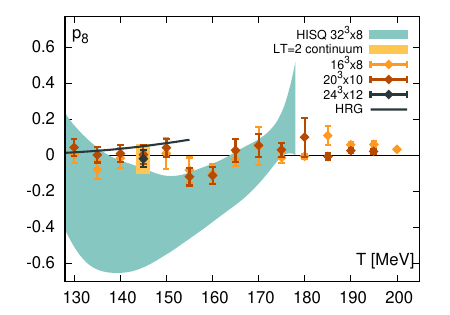}
    \end{center}
    \vspace{-0.6cm}
    \caption{Our lattice results for the strangeness neutral fluctuations $p_6$ (top) and $p_8$ (bottom). The HISQ (splined) results are shown as light blue bands. Our new results for $p_6$ are shown as colored points for finite lattice spacing, and as a yellow band for the continuum. For $p_8^B$, our $N_\tau=8,10$ results are shown in different colors, together with the HISQ (splined) results.  
    \label{fig:p6p8}        
}
\end{figure}

In the main text we presented results 
for zero strangeness chemical potential 
$\mu_S=0$. A phenomenologically more 
relevant case is that
of strangeness neutrality, where the strangeness
chemical potential has to be tuned as a function
of the baryochemical potential 
$\mu_S(\mu_B)$ such that the expectation 
value for the strangeness density is zero $n_S \propto \frac{\partial p}{\partial \mu_S} = 0$. This is
done to every order in the Taylor expansion
in $\mu_B$. The strangeness chemical potential
itself is expanded as follows:
\begin{equation}
\begin{aligned}
\mu_S(\mu_B,T) = s_1 (T) \left( \frac{\mu_B}{T} \right) 
&+ s_3 (T) \left( \frac{\mu_B}{T} \right)^3  \\
&+ s_5 (T) \left( \frac{\mu_B}{T} \right)^5  
+ \dots
\end{aligned}
\end{equation}
where the coefficients $s_1,s_3$ and $s_5$ are determined by solving $\frac{\partial p}{\partial \mu_S}=0$ order-by-order in the
expansion parameter $\frac{\mu_B}{T}$. Our 
results for these coefficients are shown 
in Fig.~\ref{fig:s1s3s5}. We also show the 
HRG predictions for these coefficients. 
We observe a good agreement for
temperatures below $145$MeV. 

Once these coefficients
are known, we can also calculate the 
Taylor coefficients of the 
pressure on the strangeness
neutral line:
\begin{equation}
    p_n \equiv \left( \frac{\partial ^n (p/T^4)}{\partial (\mu_B/T) ^n} \right)_{n_S =0} \rm.
\end{equation}
Our results for the Taylor coefficient 
ratios $p_4/p_2$, $p_6/p_2$ and $p_8/p_2$ 
are shown in 
Fig.~\ref{fig:p42p62p82}. We observe 
mild cut-off effects. Again, the HRG 
predictions are also shown. 

We also show our results for $p_6$ 
and $p_8$ in Fig.~\ref{fig:p6p8}. 
We show both the finite spacing results 
$16^3 \times 8$, $20^3 \times 10$ and 
$24^3 \times 12$, as well
as the continuum extrapolated results.
Like in the main text, for the eith order
coefficients $p_8$ the $24^3 \times 12$ result,
as well as the continuum extrapolated result is only available at a single temperature of $T=145$MeV. Our results are compared to results 
of the HotQCD collabotion using $32^3 \times 8$
HISQ lattices from Ref.~\cite{Bollweg:2022rps}.
All conclusions are similar to the case of the coefficients $\chi^B_6$ and $\chi^B_8$, with the
exception that unlike the case of $\chi^B_6$ and $\chi^B_8$, for the coefficients 
of $p_6$ and $p_8$ there are no results
from the imaginary chemical potential 
method using the 4stout $48^3 \times 12$ 
lattice action.

\subsection*{\label{stat}Statistics, parameters}

\begin{table}[h!]
\begin{center}
\begin{tabular}{|c||c|c|c|}
\hline
$T$ [MeV] & $16^3\times8$ & $20^3\times10 $ & $24^3\times12$ \\
\hline
130&31741   &71090  & 68689 \\
135&33528   &106403 & 66960 \\
140&34977   &69690  & 75229 \\
145&336975  &188571 & 111435 \\
150&65374   &108481 & 81590 \\
155&34057   &96985  & 89559 \\
160&37145   &68619  & 94053 \\
165&156044  &67668  & 98744 \\
170&34397   &42314  & 11831 \\
175&34180   &36522  & 12089 \\
180&30594   &25229  & 12727 \\
185&30951   &18396  & 13066 \\
190&30293   &18267  & 7141  \\
195&31276   &15008  & 7199  \\
200&31919   &13346  & 7390  \\
\hline
\end{tabular}
\end{center}
\caption{\label{tab:stat}
Number of configurations analyzed on our three lattice geometries.
}
\end{table}

\bibliography{thermo}

\end{document}